\documentclass[prx,onecolumn,showpacs,amsmath,amssymb,superscriptaddress]{revtex4-2}
\usepackage{graphicx}
\usepackage{amsmath}
\usepackage{amssymb}
\usepackage{dcolumn}
\usepackage{latexsym}
\usepackage{rotating}
\usepackage{booktabs}
\usepackage[dvipsnames]{xcolor}
\usepackage{float}
\usepackage{epsfig}
\usepackage{psfrag}
\usepackage{natbib}
\usepackage{bm}
\usepackage{eucal}
\usepackage{mathrsfs}
\usepackage{braket}
\usepackage{enumerate}
\usepackage{longtable}
\usepackage{wrapfig}
\usepackage{bm}
\usepackage{hyperref}
\usepackage{amsfonts}
\usepackage{dcolumn}
\usepackage{bm}
\usepackage{xspace}
\usepackage{color, soul}
\usepackage[normalem]{ulem}
\usepackage{multirow}

\newcommand{\be}{\begin{equation}}
\newcommand{\ee}{\end{equation}}
\newcommand{\bn}{\begin{eqnarray}}
\newcommand{\en}{\end{eqnarray}}

\usepackage[version=4]{mhchem}

\newcommand{\qsgwh}{QS$G\hat{W}$\xspace}

\def\x2y2{{x^2-y^2}}

\definecolor{seagreen}{RGB}{20,150,75}

\usepackage{color} 

\usepackage{hyperref}
\hypersetup{
colorlinks=true,final=true,
        linkcolor=red,
        citecolor=blue,
        filecolor=blue,
        urlcolor=blue,
}

\graphicspath{{figs/}}
\begin{document}

\title{Quasiparticle Description of Angle-Resolved Photoemission Spectroscopy for \ce{SrCuO2}}
\author{Dimitar Pashov}
\affiliation{Theory and Simulation of Condensed Matter, King's College London, The Strand, London WC2R2LS, UK}
\author{Casey Eichstaedt}
\affiliation{Materials, Chemical and Computational Science Directorate, National Renewable Energy Laboratory, 15013 Denver West Parkway, Golden CO 80401}
\author{Swagata Acharya}
\affiliation{Materials, Chemical and Computational Science Directorate, National Renewable Energy Laboratory, 15013 Denver West Parkway, Golden CO 80401}
\author{Mark van Schilfgaarde}
\affiliation{Materials, Chemical and Computational Science Directorate, National Renewable Energy Laboratory, 15013 Denver West Parkway, Golden CO 80401}

\begin{abstract}

\ce{SrCuO2} has long been considered a near-archetypal realization of a quasi-one-dimensional (1D) system of interacting
electrons with short-range interactions. Within this framework, experimental observations---interpreted through the lens
of the 1D Hubbard model---suggest that electron and hole excitations decay into two types of (unphysical) collective
bosonic modes: ``spinons'', which carry the spin degree of freedom, and ``holons, which carry the charge degree of
freedom. This model, known as spin-charge separation, is most directly evidenced by angle-resolved photoemission
spectroscopy (ARPES), where a photo-induced hole decays into a continuum of these excitations.  Here we present an
alternative perspective grounded in first-principles, self-consistent, and parameter-free many-body perturbation
theory. In this revised quasiparticle framework, ARPES can be understood as a one-body effect arising from mild disorder
in a long-range antiferromagnetic ground state.  the emergence of the so-called spinon branch arises naturally from spin
disorder, the anomalous linewidths are accurately captured, and we provide a compelling explanation for the spectral
weight observed at the non-magnetic zone boundary. This reinterpretation provides a unified explanation for key
experimental signatures previously attributed to spin-charge separation, including features observed in optical
conductivity.  Additionally, we show that \ce{SrCuO2} exhibits a nontrivial interchain coupling that significantly
influences both its one-particle and two-particle spectral functions. By comparing the spectral features of \ce{SrCuO2}
with those of \ce{La2CuO4}, we argue that \ce{SrCuO2} shares notable similarities with the two-dimensional
cuprates---both being rooted in a common \ce{CuO4} plaquette-based molecular orbital framework.


\end{abstract}
\maketitle


\section{Introduction}

The one-dimensional (1D) cuprates---exemplified by \ce{SrCuO2}--- are widely regarded as prototypical realizations of 1D
Mott insulators. A hallmark feature frequently cited in the literature is spin-charge
separation~\cite{voit-scs,nozieres99,giamarchi2004quantum,KIM2001}, a phenomenon predicted by the 1D Hubbard
model~\cite{Lieb-Wu} in which electron and hole excitations fractionalize into two distinct bosonic collective modes:
spinons, which carry spin but no charge~\cite{kivelson87,zou88}, and holons, which carry charge but no spin. Among the
most direct experimental signatures of this many-body effect---beyond the scope of any one-particle description---is
provided by angle-resolved photoemission spectroscopy (ARPES). In this framework, the electron removal operator
factorizes into spinon and holon components, such that the resulting spectral weight reflects not quasiparticle
dispersions but the energy and momentum conservation laws governing these fractionalized excitations. This
interpretation has been invoked to explain the presence of an anomalous low-energy peak in the ARPES spectrum---one that
lies beyond the predictions of conventional band theory~\cite{Kim96,Kim97,Shen06}, as the the theory of spin-charge
separation lies outside a one-particle picture.  Note that in various other solid state oxides a similar argument in
favor of strong many-body correlations has been made; see e.g. Fig. 2 in Fujimori et
al.~\cite{PhysRevLett.69.1796}. Often in the literature, observation of spectral weight in places (in energy and
momentum) where bands computed from density functional theory
(DFT) are absent is considered as a signature of strong many-body interactions.  However, this canonical definition of
correlation is arbitrary, not only because DFT is only one possible choice of single-particle electronic structure, but
also because symmetry can be explicitly broken through intersite fluctuations and still be described at the mean-field
level. Oxides can have instantaneous structural disordering that reduces the local symmetry while keeping the global
symmetry intact and that can strongly influence the low-energy states around Fermi level and band
gap~\cite{Zunger23}. In systems with partially filled \textit{d}-states, the spin degree of freedom in the electronic
structure can be another source of disordering. The physics underlying spin fluctuations leading to spectral weight
redistribution is even more involved and interesting. When spin fluctuations are within a single site, atomic multiplets
can appear which entail a strictly a multi-Slater-determinant scenario beyond DFT or any other quasiparticle
theory. On-site scenarios have traditionally been described by ligand field theory, or by dynamical mean field
theory. On the other hand, intersite spin fluctuations and their impact on electronic structure can be reasonably
captured within a quasiparticle theory~\cite{Zunger21,Zunger23,Watson24}.

Here, we demonstrate that the measured ARPES spectrum can be understood within a conventional quasiparticle
framework. Features commonly attributed to holons arise instead from the influence of \textit{long-range}
antiferromagnetic correlations, while the additional spectral branch often ascribed to spinons can be explained by the
presence of spin disorder within this ordered background. Crucially, our interpretation remains entirely within the
bounds of conventional band theory and does not invoke the fractionalization of the electron into formally unphysical
collective excitations and is able to capture features in the photoemission spectrum that are absent in the 1D Hubbard
model.  We also demonstrate that the first-order electronic structures of \ce{SrCuO2} and the 2D cuprates---represented
here by \ce{La2CuO4}---are remarkably similar. While Anderson~\cite{anderson90} previously proposed a related
connection, he interpreted the 2D Hubbard model in the normal metallic state---commonly accepted as a minimal model for
the 2D cuprate superconductors---as exhibiting quasi-1D Luttinger liquid behavior. In contrast, we argue that both
systems can be more accurately described within a conventional band theory framework. Specifically, we show that their
low-energy electronic structures are well captured by a ligand field picture, in which copper resides in a $d^{9}$
configuration coordinated by four oxygen atoms arranged in a planar geometry for the 2D cuprates and in a linear chain
for the 1D compounds (Fig.~\ref{fig:Fig1}(c)).

In both cases, adjacent Cu neighbors share two of the four O atoms.  That being said, a molecular unit within a
\ce{CuO4} plaquette is not sufficient for the optical response. We show that for both \ce{SrCuO2} and \ce{La2CuO4}, the
important electron-hole excitations at low energies involve intersite Cu-Cu $d$-$d$ transitions. \ce{SrCuO2} and
\ce{La2CuO4} share much in common, but important differences also appear.  We show here that the low-lying excitons in
\ce{SrCuO2} are highly anisotropic and stretch out more in one direction (along the chain) while in \ce{La2CuO4} their
pattern much more reflects a 2D geometry involving several dozen molecular units. In this context, we establish the fact
rigorously that the low-energy electron-hole two-particle (excitonic) transitions in both \ce{La2CuO4} and \ce{SrCuO2}
involve intersite $d_{x^{2}-y^{2}}$ electronic orbitals in strong contrast to the onsite $t_{2g}$-$e_{g}$ transitions
prescribed by the $d^{9}$ Sugano-Tanabe diagrams~\cite{sugano1954,suganobook} which are grounded in ligand-field theory
or that described via orbital waves (``orbitons'')~\cite{Saitoh2001} to interpret the seminal $L_3$ edge resonant
inelastic X-ray scattering (RIXS) data for \ce{Sr2CuO3}~\cite{vandenbrink2013,Schlappa}.

{\color{black}
\section{Results}
}

{\color{black}
\subsection*{Theoretical framework}
}
\color{black}

To rigorously describe the electronic structure of \ce{SrCuO2} and \ce{La2CuO4} from first principles, we employ a
self-consistent implementation of many-body perturbation theory (MBPT), with excitonic effects included
\textcolor{black}{in a self-consistent manner with addition of ladder diagrams in both the polarizability and
  self-energy~\cite{Cunningham23}, built into the Questaal package~\cite{questaal_paper,questaal_web}.}
\textcolor{black}{
A distinguishing feature of Questaal is its quasiparticle self-consistent form of \textit{GW},
QS\textit{GW}. \textit{GW} is usually implemented as an extension to density-functional theory (DFT), i.e. \textit{G}
and \textit{W} are generated from DFT.  QS\textit{GW}~\cite{Faleev04} may be thought of as an optimized form of the
\textit{GW} approximation of Hedin~\cite{mark06qsgw,Kotani07,Beigi17}.  Self-consistency removes dependence on the
starting point~\cite{mark06qsgw,Kotani07} and also makes it possible to more systematically and reliably predict
physical observables, especially those sensitive to self-consistency such as the magnetic moment~\cite{Faleev04} and
response functions~\cite{Acharya21a}.}

\textcolor{black}{
However, QS\textit{GW} systematically overestimates bandgaps by 10-20$\%$~\cite{Shishkin07}. The primary source of this
discrepancy is the missing electron-hole (excitonic) correlations. We recently built a self-consistent extension of
QS\textit{GW}, called QS$G\hat{W}$, where the screened Coulomb interaction \textit{W} includes electron-hole
interactions by solving a Bethe-Salpeter equation~\cite{Cunningham23}.  Crucially, \textit{G}, $\Sigma$ and $\hat{W}$
are iterated until they mutually converge. This contrasts with single-shot \textit{GW} approaches that do not include
changes in electronic structure from changes in the density or contributions from electron-hole attraction, thus
biased by unreliable eigenvalues and eigenfunctions.
}

This formalism has consistently yielded high-fidelity results for both one-particle
and two-particle observables across a broad class of insulating materials, including \ce{La2CuO4}, for which extensive
experimental data are available~\cite{Imada98,Norman03}. As validation, we benchmark theoretical spectral functions
against ARPES and compare computed optical conductivities $\sigma (\omega)$ with ellipsometry data. Where experimental
results exist, our theory achieves an agreement ranging from good to excellent.  Importantly, this framework enables a
physical interpretation of experimental features without resorting to exotic phenomena~\cite{Mazin2022}. In particular,
we demonstrate that the long-range spin disorder captured within a high-level quasiparticle theory---built from Feynman
diagrammatic expansions---is sufficient to account for the spectral feature commonly identified as the spinon
branch. Rather than requiring a breakdown of the electron into spin and charge components, this feature naturally
emerges from the interplay between extended electron and hole wavefunctions and the long-range antiferromagnetic
correlations intrinsic to \ce{SrCuO2}. Just as Peierls instabilities arise from electron-phonon interactions in 1D
systems, here the cooperative effects between spin disorder and delocalized charge carriers give rise to a band that can
be accurately described within a single-Slater-determinant framework derived from MBPT. Within this framework, this
low-energy feature persists in our second Brillouin zone (BZ) (the second half for the nonmagnetic BZ) which was
attributed to a background effect~\cite{Kim97}.  Additionally, the broad linewidth observed in ARPES can be rationalized
by accounting for disorder perpendicular to the \ce{CuO2} chains. Finally, our optical data has its origin in rigorous
conventional electronic structure theory and does not invoke electron-hole excitations comprised of spinless
``holon-doublon'' or ``holon-antiholon'' pairs~\cite{Jeckelmann00,essler01,Jeckelmann03,Kim04,Jeckelmann07}.

\subsection*{Lattice and Magnetic Structure of SrCuO$_2$}

\begin{figure}[t]
\centering
\includegraphics[height=7.0cm]{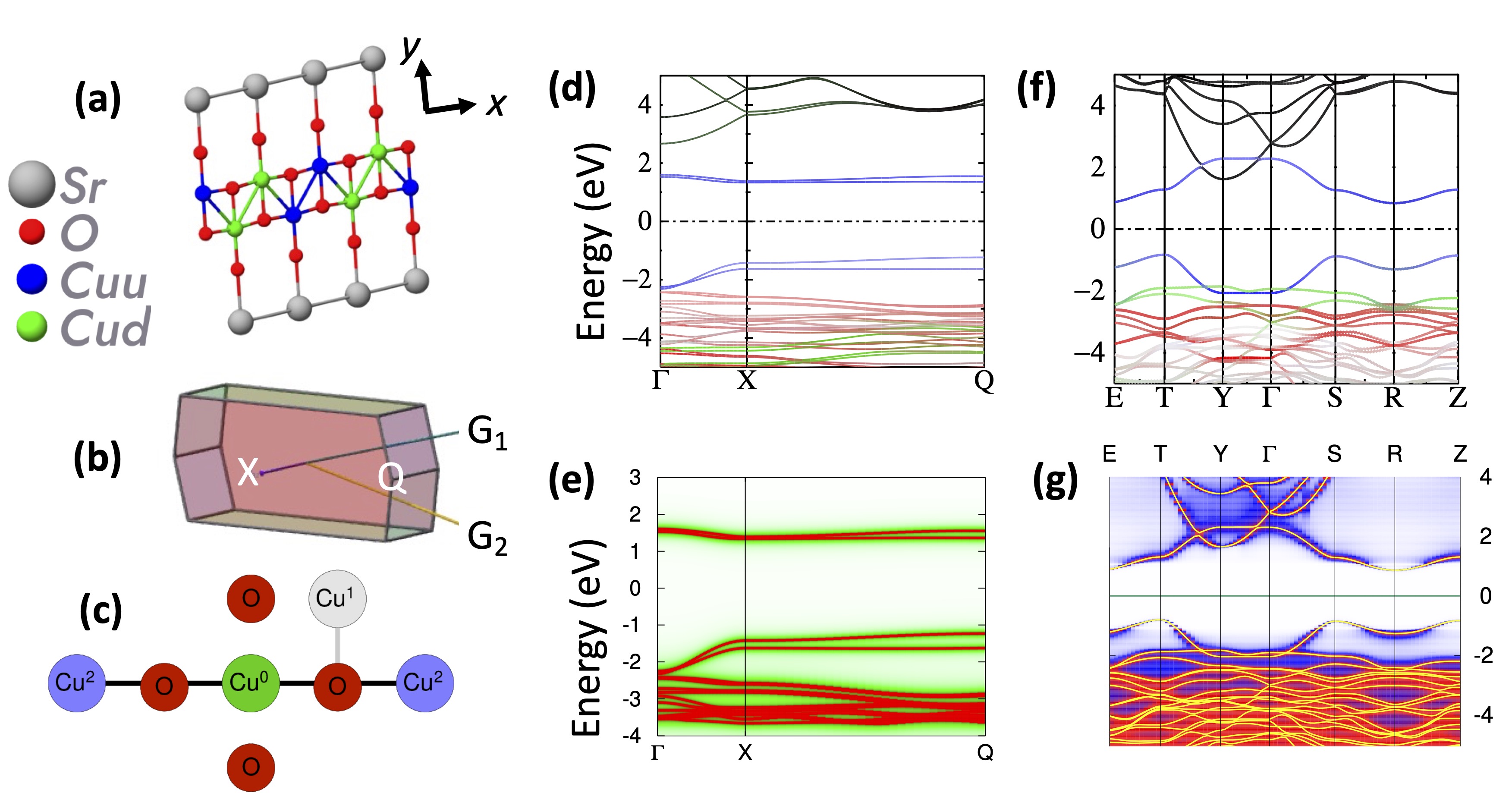}
\caption{ (\textbf{a}) \ce{SrCuO2} in the minimal (16 atom) unit cell.  Two unit cells are shown.  Cu, O, and Sr almost
  lie a single plane.  (\textbf{b}) corresponding (minimal) Brillouin zone with C2/$m$ symmetry.  X is a zone-boundary
  point along the chain, at (0,0,1/2), i.e. the midpoint of the reciprocal lattice vector along \textit{x}.  Q, which
  corresponds to (5/16,5/16,1/2), is where the valence band maximum falls, according to QS$G\hat{W}$ theory.
  (\textbf{c}) Molecular fragment for \ce{SrCuO2} and \ce{La2CuO4}. The O atoms form a nearly perfect square.  For
  both systems, Cu $d_{x^2-y^2}$ and O $p_{x}$ combine to form strong $\sigma$ bonds and antibonds in the chain,
  depicted by the bar.  Cu$^{(1)}$ is not present in \ce{La2CuO4}; instead Cu$^{(2)}$ appears on both \textit{x}
  and \textit{y} axes.  In \ce{SrCuO2} Cu$^{(1)}$ is present but weakly bonds to Cu$^{(0)}$ and Cu$^{(2)}$ on the chain,
  as explained in the text.  (\textbf{d}) Noninteracting energy bands of \ce{SrCuO2} within the quasiparticle
  self-consistent $G\hat{W}$ approximation, for the unit cell of panel (\textbf{a}).  The zero defines the Fermi level
  $E_{F}$, placed at midgap.  Colors depict orbital weights: blue, red, green project onto Cu $d_{x^2-y^2}$, $t_{2g}$,
  $d_{z^2}$ character, respectively.  Black depicts Sr content.  O content is depicted by bleaching out the colors.
  States between $E_{F}-4$\,eV and $E_{F}+2$\,eV are composed of roughly equal admixtures of Cu $3d$ and O $2p$
  character, while states above $E_{F}+2$\,eV consist of mostly Sr character (La character in \ce{La2CuO4}).
  (\textbf{e}) Interacting band structure calculated from the dynamical QS$G\hat{W}$ self-energy $\Sigma(q,\omega)$, in
  the window ($-$4,2.5)\,eV.  (\textbf{f}) Noninteracting energy band structure of \ce{La2CuO4}, with the same
  color scheme as panel (\textbf{d}).  (\textbf{g}) Corresponding interacting band structure of \ce{La2CuO4}.  }
\label{fig:Fig1}
\end{figure}

The unit cell of \ce{SrCuO2}~\cite{SrCuO2-crystal} consists of puckered, zigzag chains of Cu atoms running along the
\textit{x}-axis (our chosen convention), each decorated by oxygen atoms to form a square planar geometry. Each Cu ion is
coordinated by four nearest-neighbor O atoms lying approximately in a plane, with Cu–O bond lengths of
{$1.94{\pm}0.03$\,\AA} and O-Cu-O bond angles $90{\pm}2.5^{\circ}$ (Fig.~\ref{fig:Fig1}(c)). Strontium atoms reside
between the chains, acting as charge reservoirs but contributing minimally to the low-energy electronic structure.
It is helpful to conceptualize the structure as comprising pairs of magnetically weakly-coupled \ce{CuO2} linear
chains. The minimal magnetic unit cell we adopt contains two such antiferromagnetically ordered \ce{CuO2} chains,
necessitating a doubling of the smallest structural repeat along \textit{x} to accommodate alternating spin order
between Cu$^{(0)}$ and Cu$^{(2)}$ atoms. We refer to this 16-atom unit as the ``minimal cell,'' and construct the full
3D crystal by periodically replicating it along the \textit{y}- and \textit{z}-directions.
Along \textit{y}, chains are widely spaced, resulting in negligible interchain coupling---spectral features show little
dependence on magnetic order along this direction. In contrast, along \textit{z}, Cu ions are separated by 3.58\,\AA ,
roughly 10\% shorter than the Cu$^{(0)}$–Cu$^{(2)}$ distance along \textit{x}. While interlayer coupling along
\textit{z} is weaker due to the absence of mediating oxygen atoms, it is not negligible. As we will show, variations in
magnetic order along \textit{z} contribute to the broad linewidths observed in ARPES, which cannot be explained by the
sharply-peaked predictions of the 1D Hubbard model~\cite{Shen06}.
For comparison, in \ce{La2CuO4}, Cu ions have a nearly identical ligand environment, with Cu–O bond lengths of
{$1.87$\,\AA} and bond angles of $90{\pm}1^{\circ}$ . Fig.~\ref{fig:Fig1}(c) illustrates the fundamental \ce{CuO4}
molecular units for both \ce{SrCuO2} and \ce{La2CuO4}. In both materials, a Mott gap arises from bonding–antibonding
splitting of $e_g$ states between neighboring Cu atoms mediated by bridging oxygen: the bonding state is filled, while
the antibonding state remains empty. Cu$^{(0)}$ also weakly couples to Cu$^{(1)}$ via $\pi$-bonding through oxygen,
although this interaction is relatively small, as quantified in the electronic structure analysis. This bonding motif is
shared by both compounds.
Although the overall electronic structure of \ce{SrCuO2} and \ce{La2CuO4} is remarkably similar, structural differences
introduce important secondary modifications. Even among 2D cuprates, such variations significantly impact the electronic
structure---for instance, compare \ce{Nd2CuO4}, which lacks apical oxygen, with \ce{La2CuO4}, which possesses
it~\cite{PhysRevB.82.060513,PhysRevLett.69.1796,weber2012scaling,Acharya18}.

Our QS$G\hat{W}$ calculations reveal that \ce{SrCuO2} is not truly 1D, but rather a highly anisotropic 2D system, akin
to the quasi-1D antiferromagnet CrSBr. In the sections that follow, we elaborate on the parallels and distinctions
between \ce{SrCuO2} and \ce{La2CuO4} and explore their implications for correlated-electron physics.

\subsection*{Electronic Structure of SrCuO$_2$ and its relation to La$_2$CuO$_4$}

{\color{black}
We start with the minimal structure: a 16-atom unit cell with structural information given in Table 1 in the
Supplementary Information.  We will refer to the Brillouin zone corresponding to this structure as the ``minimal
Brillouin zone'' (mBZ).  This cell consists of a single chain; every replica along \textit{z} and \textit{y} is aligned
ferromagnetically to this one.

Fig.~\ref{fig:Fig1}(c) shows the energy band structure for this cell, calculated in the QS$G\hat{W}$
approximation.  The two highest occupied states and lowest unoccupied states consist mainly a mixture of Cu $d_{3z^2{-}1}$
and O $p$ orbitals.  These four states correspond to bonding and antibonding states of the four Cu in the unit cell.
Dispersion of these states is much larger on G-X line of the Brillouin zone (excursions in (0,0,$k_{x}$) than the X-Q
line (excursions normal to $k_{x}$), because the orbitals comprising these states are oriented along \textit{x}.
Nevertheless some dispersion can be seen on the X-Q line as well; this will be significant for the photoemission
discussed next.  The situation is opposite for states of Cu $d_{x^2-y^2,xy}$ character between $-$2.5 and $-$3\,eV,
depicted as pastel red (the color is faded because of the strong admixing from O).  According to the theory, the valence band
maximum falls at Q.  Panel (d) shows the interacting band structure (spectral function), computed from the QS$G\hat{W}$
dynamical self-energy.  Many-body broadening of the quasiparticle peaks is small compared to other cuprates; compare for
example to CuO and \ce{La2CuO4} (Figs 25, 26, 28 in Ref.~\cite{Cunningham23}), which suggests \ce{SrCuO2} is not
strongly correlated.
}

\ce{SrCuO2} is most easily compared to \ce{La2CuO4} using the 16 atom minimal cell.  This is a third of the size of the
experimentally reported 3D-ordered unit cell in the low-temperature ground state, where chains are stacked antiferromagnetically
(AFM) along \textit{y} and \textit{z}~\cite{aforder}. By ordering antiferromagnetically along the \textit{z} axis,  we denote it as the ``standard
cell,'' and employ `mBZ' to denote the Brillouin of the minimal cell (Fig.~\ref{fig:Fig1}(b)), keeping in mind the BZ of
the standard cell is half this size. It is noteworthy to mention that magnetic ordering along \textit{y} does not affect any of the properties. 
It is also important to remember that ARPES studies use the convention of the non-magnetic BZ as the
true one, which is twice the size of our mBZ.  To compare with ARPES, we will use a Brillouin zone unfolding technique to represent energy bands of more
complex cells in terms of the mBZ where spin-charge separation is claimed to be most prominent.  The unfolding procedure can be found in Ref.~\cite{TiSe2}.
Figs.~\ref{fig:Fig1}(d,e) and Figs.~\ref{fig:Fig1}(f,g) show the quasiparticle band structure and interacting spectral function for the minimal cell in \ce{SrCuO2} and
the 14 atom cell in \ce{La2CuO4}.  In a molecular picture a pair of states consisting of adjacent
Cu-Cu $d_{x^2-y^2}$ orbitals form bond-antibond pairs to form the LUMO and HOMO (blue bands).
Deeper in the valence are states of $t_{2g}$ character (red), and of $d_{z^2}$ character (green).  For both \ce{SrCuO2}
and \ce{La2CuO4}, Cu $d_{x^2-y^2}$ and O $p$ contribute roughly equally to the (blue) frontier orbitals of
Fig.~\ref{fig:Fig1} (the O content is depicted by bleaching out the colors).  QS$G\hat{W}$ stabilizes a local Cu
magnetic moment in both, 0.58\,$\mu_B$ in \ce{SrCuO2} and 0.53\,$\mu_B$ in \ce{La2CuO4}, and in both cases the
local moment is necessary for the band gap opening; without it, the system becomes metallic.  The reason can be
understood as follows.  Adjacent Cu $d_{x^2-y^2}$ orbitals couple though O $p_{x}$ to form $\sigma$ bonds,
Fig.~\ref{fig:Fig1}(c), in symmetric (bonding) or antisymmetric (antibonding) combinations.  Square-planar  crystal fields cause
$e_{g}$ to split, leaving only $d_{x^2-y^2}$ near the gap. Cu is in a $d^{9}$ configuration, so that there are two
electrons per Cu pair to occupy the pair of $d_{x^2-y^2}$ orbitals on the two atoms.  If the system is nonmagnetic, the
$\uparrow$ and $\downarrow$ states are degenerate, so two electrons occupy four orbitals, making the system metallic.
However, a local moment forms and spin-splits $d_{x^2-y^2}$, leaving only one orbital available per atom and spin.  Then for
each spin, the bonding states are occupied and the antibonding states are empty, forming a gap.  The picture for \ce{SrCuO2} is slightly
complicated by the weak coupling between the parallel chains: it slightly splits both the conduction and valence
frontier orbitals (blue bands, Fig.~\ref{fig:Fig1}(c)).  Indeed, a measure of the interchain to intrachain coupling is
the ratio of this splitting to the gap.


Moving to the Bloch picture, the HOMO, LUMO, and other states broaden into bands.  For \ce{SrCuO2}, dispersion is
strongest along the chain axis ($\Gamma$-X line) where the hopping matrix elements are large.  The QS$G\hat{W}$
bandwidth is 0.82\,eV for the minimal cell, slightly smaller than the dispersion observed for the holon band in
photoemission~\cite{Shen06}.

\textit{Correlations:} \ce{SrCuO2} and \ce{La2CuO4} show striking differences in the degree of correlation detected
in QS$G\hat{W}$ theory.  Figs.~\ref{fig:Fig1}(e,g) show the interacting band structure correlative to the noninteracting
bands in Figs.~\ref{fig:Fig1}(d,f).  They are computed from the spectral function generated by the dynamical self-energy
$\Sigma(\mathbf{q},\omega)$ as distinct from the static quasiparticlized $\Sigma^0(\mathbf{q})$.  (By the QS$G\hat{W}$
construction, the poles of the noninteracting $G^{0}$ and the interacting $G$ align, so the dynamical $\Sigma$ causes no
shift in the spectral peaks, bandgap or bandwidth, in contrast to non self-consistent MBPT.)
Many-body broadening of the quasiparticle peaks is small compared to other cuprates; compare for example to CuO and
\ce{La2CuO4} (Figs 25, 26, 28 in Ref.~\cite{Cunningham23}), which suggests \ce{SrCuO2} is well described by a single Slater determinant in the undoped case.  On the other hand, correlations from disorder (extrinsic defects, spin disorder) may be greater in \ce{SrCuO2}, because the bandwidths are probably narrower on average and scattering is channeled mostly along one dimension.


{\color{black}
\subsection*{Spin Fluctuations}
}

\begin{figure}[t]
\centering
\includegraphics[height=5.0cm]{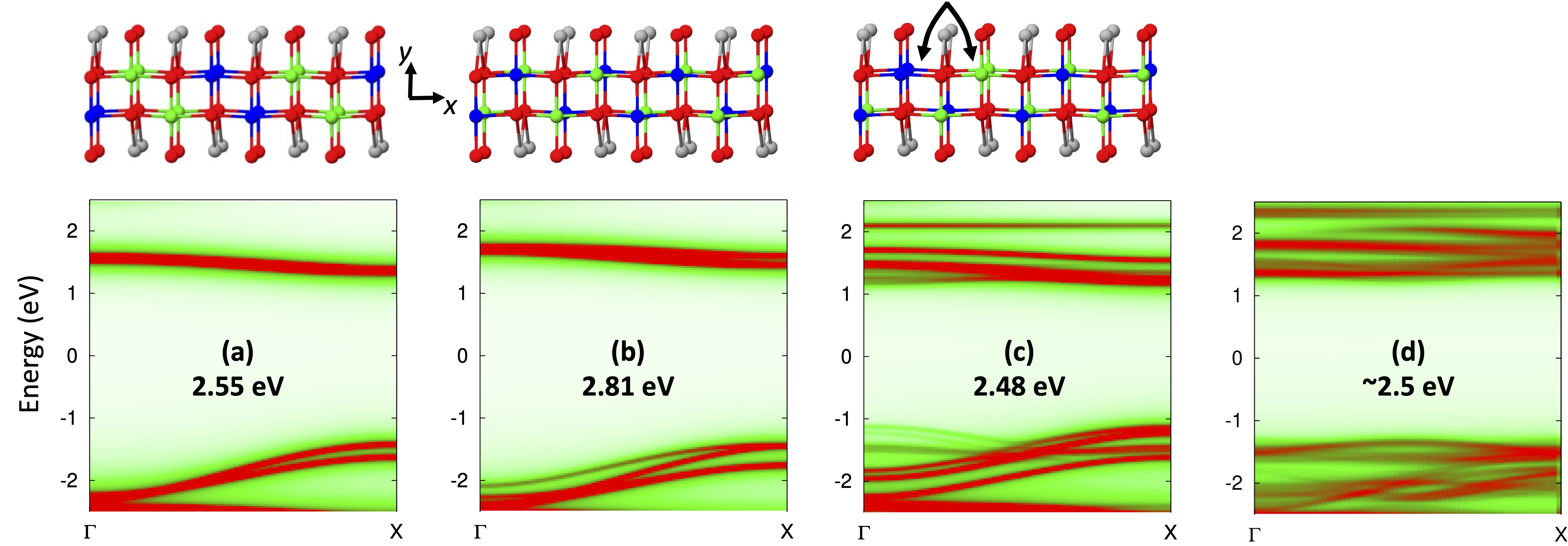}
\caption{Spectral functions for selected spin orderings of a 64-atom superlattice, unfolded to the minimal Brillouin zone, on the $\Gamma$-X line.
The bandgap is given for each panel.
\textbf{(a)} a trivial supercell of the mBZ (equivalent to bands of Fig.~\ref{fig:Fig1}(c)).
  \textbf{(b)} spins in alternate planes along \textit{z} are spin flipped (standard cell)
  \textbf{(c)} Permutation of \textbf{(b)} where a pair of adjacent Cu marked by the arrows are spin flipped.
  \textbf{(d)} Configurationally averaged spectral functions from 20 random spin configurations.\color{black}
 }
\label{fig:Fig2}
\end{figure}

Scattering from spin fluctuations is not taken into account, as QS$G\hat{W}$ has no diagram beyond the Fock exchange in
the spin channel.  Zunger and co-workers incorporated spin fluctuations in a static mean-field approximation by
constructing supercells of the minimal cell with disordered spin configurations.  They employed this technique for
FeSe~\cite{Zunger21} in a DFT context, with reasonable results.  We have employed it for CrSBr in a QS$G\hat{W}$
framework, to explain the observed change seen in photoemission from AFM$\rightarrow$PM phases~\cite{Watson24}.  We
employ the same approach here to approximately account for spin fluctuations.
We present results for 64-atom cells, superlattices that double the
\textit{z} axis along the chain, and elongate the \textit{z} axis to include two planes instead of one (see cartoon in
Fig.~\ref{fig:Fig2}(a)).

If \ce{SrCuO2} were truly 1D, the spectral function would not depend on the relative
spin ordering of adjacent chains.  This is indeed the case for the \textit{y} axis where chains are widely spaced;
however, the transition FM$\rightarrow$AFM ordering in \textit{z} (minimal cell $\rightarrow$ standard cell) causes
noticeable differences: the bandwidth of the holon band increases, the bandgap widens by $\sim$250\,meV, and a new band
appears on the $\Gamma$-X line, with reduced QP weight (compare Figs.~\ref{fig:Fig2}(a) and (b)).  The bandgap is larger in the AFM case
because electrons are more confined: they see a higher barrier in the \textit{z} direction---an effect
also seen in the layered, quasi-1D CrSBr system, where the FM$\rightarrow$AFM gap change is 100\,meV~\cite{Menon25}.

\ce{SrCuO2} has a N\'eel temperature of $\sim$5\,K~\cite{Neelsrcuo2}, very small compared to the 317\,K reported
for carefully annealed \ce{La2CuO4}~\cite{lsco_neel}.  This is because long range order is more difficult to
sustain with reduced dimensionality, as known from the Mermin Wagner theorem.  To investigate this, we introduce
disorder into the 64-atom supercells.  For minimum disruption, we consider flipping a pair of adjacent spins in a chain
and observe how the spectral function is modified in Fig.~\ref{fig:Fig2}(c).  This figure bears a strong resemblance to
photoemission studies~\cite{Shen06}, with a new, nearly dispersionless `spinon' band emerging a few tenths of an eV below
the \color{black}{highest lying holon valence band at X}.
We modeled another limit, the fully disordered case, by selecting 20 random spin permutations
in the 64 atom supercell and configurationally averaging the spectral functions.  The resulting spectral function is
shown in Fig.~\ref{fig:Fig2}(d).  While the holon band can be made out, a pseudo-gap appears at about $-1.8$ eV and
this band splits into fragments.  The `spinon' band is also seen, but it is wrongly placed, too intense, and rather
different from the photoemission data.  A comprehensive characterization of spin disorder is beyond the scope of this
paper, but from these two limits we can conclude (1) what have been called `spinons' are ordinary quasiparticles
appearing as a consequence of spin disorder and (2) the system mostly preserves short range order, even if long range order
is absent.

\begin{figure}[t]
\centering
\includegraphics[height=9.0cm]{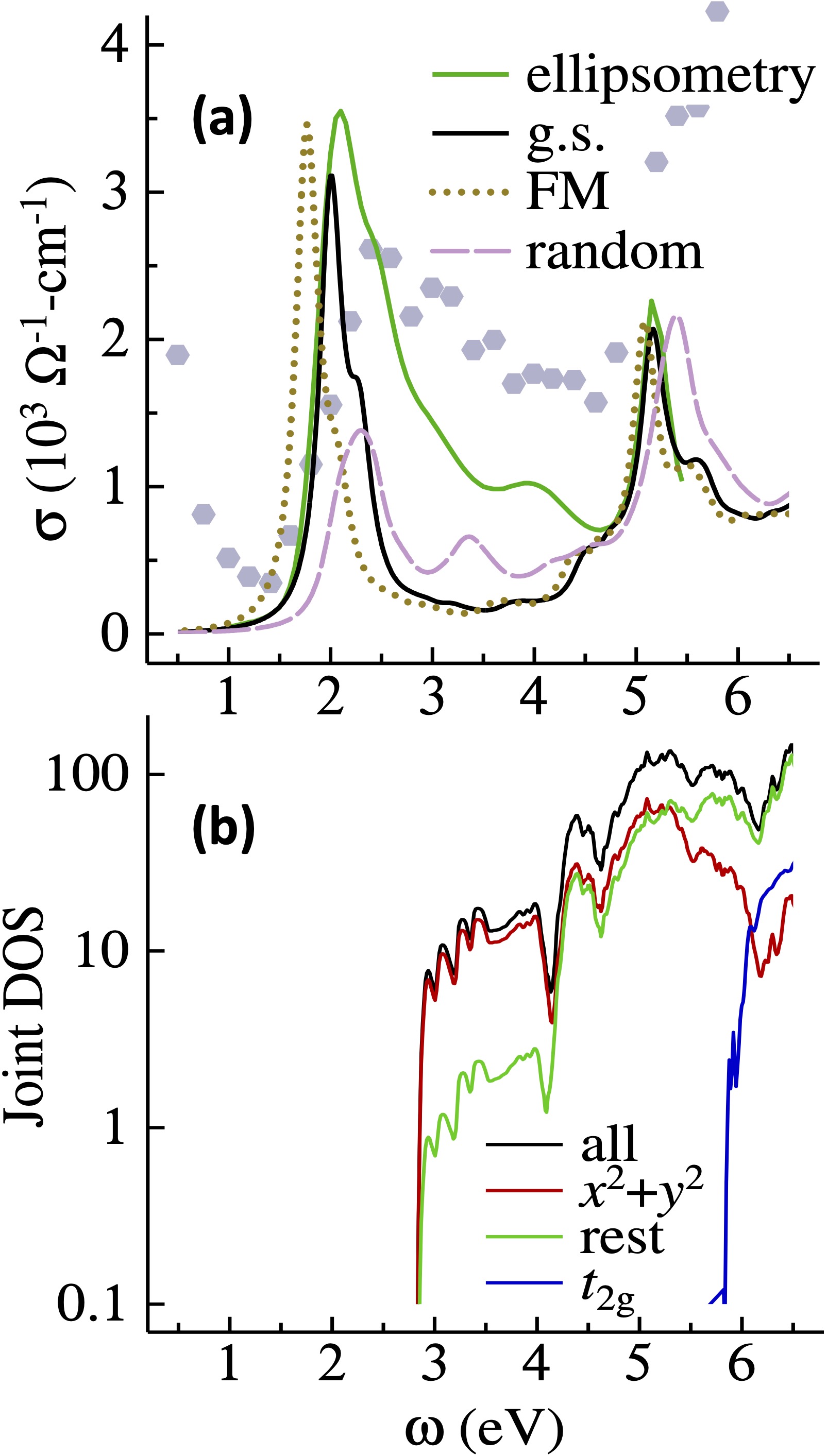}\ 
\includegraphics[height=9.0cm]{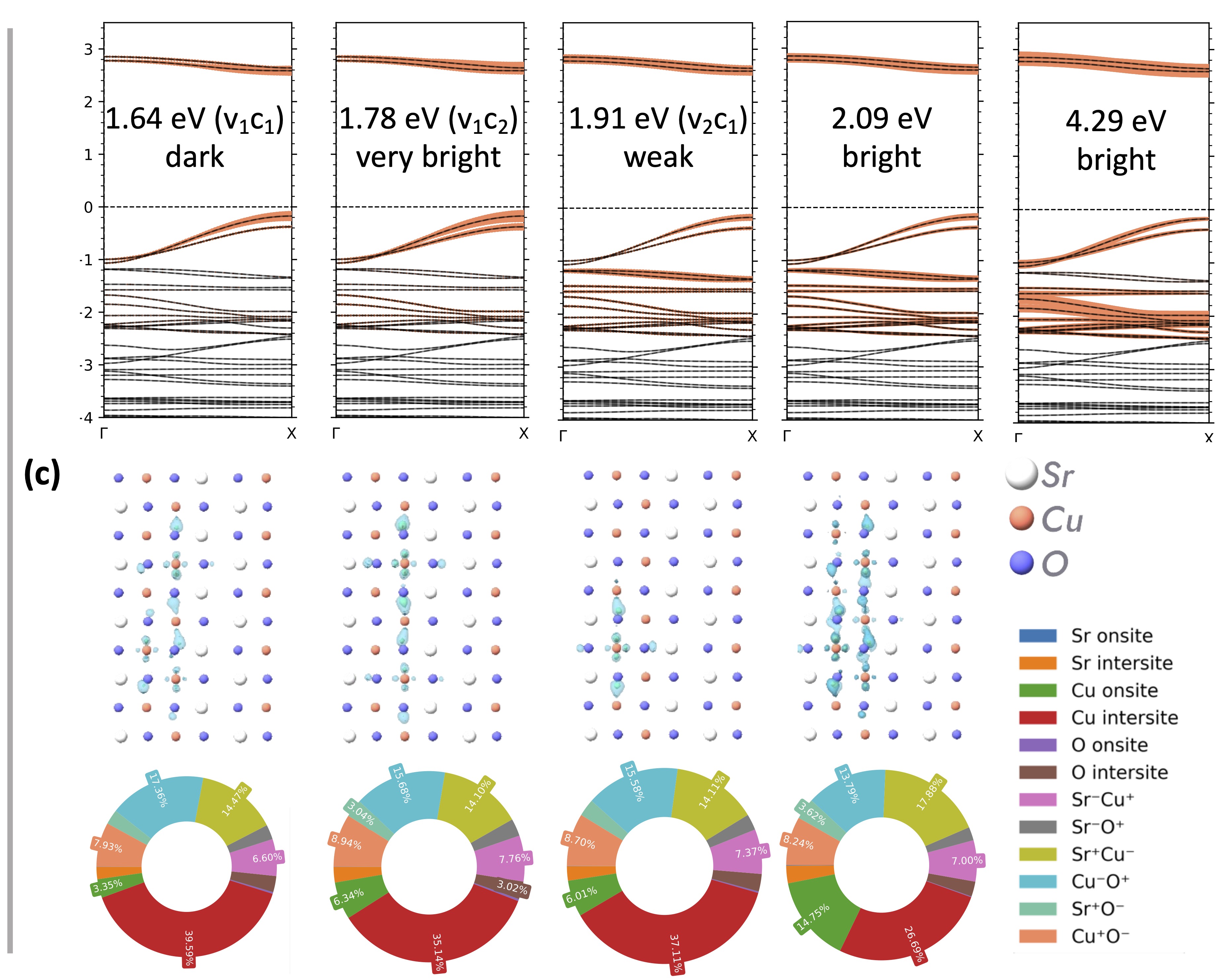}
\caption{Optical properties of  \ce{SrCuO2}.
\textbf{(a)}: conductivity $\sigma(\omega)$.
Ellipsometry data from Ref.~\cite{Hilgers09} (green);
$\sigma^\mathrm{BSE}(\omega)$ for the ground state spin configuration (black);
$\sigma^\mathrm{BSE}(\omega)$ for the minimal cell (dotted);
configurationally averaged $\sigma^\mathrm{BSE}(\omega)$ for disordered spins (dashed).
Main peak at 2.00\,eV (g.s.), 2.02\,eV (ellipsometry), 1.77\,eV (minimal), 2.29\,eV (disordered).
Also shown are RIXS spectra at $q$=0 from Ref.~\cite{Kim04} (light grey hexagons).
\textbf{(b)}: joint density-of-states.
Total (black); projection onto Cu($d_{x^2+y^2}$)\,+\,O components for both electrons and holes (red);
difference between red and black (green);
subtracting Cu($d_{x^2+y^2}$)\,+\,Cu($d_{xy}$)\,+\,O components from total (blue).
\textbf{(c)}: Selected excitons for the minimal unit cell.
\textit{Top panels:}
Energy band structure of the minimal unit cell (Fig.~\ref{fig:Fig1}), projecting the distribution of each of the three
lowest-energy excitons into band contributions (orange).  The two most strongly bound excitons (1.64\,eV and 1.78\,eV)
are composed from the highest valence frontier orbital ($v_1$) and the first or second frontier orbital ($c_1$ and
$c_2$).  Also shown are excitons that make bright peaks at 2.09\,eV and 4.29\,eV.
\textit{Middle panels:}
Real-space structure of the electron part of the exciton for a hole centered at Cu.
Atoms are denoted by white, orange and purple circles; the exciton constant-value contour
is depicted in sky blue.
\textit{Bottom panels:}
Pie chart resolving the excitons' Mulliken decomposition into atom-atom pairs, distinguishing
on-site from inter-site contributions. Superscript `+' and `$-$' refer to `electron' and `hole.'
The Cu-Cu onsite portion is small or negligible.
}
\label{fig:optics}
\end{figure}

\subsection*{Optical properties}

The optical conductivity $\sigma(\omega)$ can be calculated within the BSE from the QS$G\hat{W}$ hamiltonian, to compare
with ellipsometry measurements reported by several groups~\cite{Imada98,Popavic01,Kim04,Hilgers09}; see
Fig.~\ref{fig:optics}(a).  As they are largely similar among themselves we compare to the data in Ref~\cite{Hilgers09}
since it is the most recent and most complete.  Comparing $\sigma^\mathrm{BSE}$ (black line) to $\sigma^\mathrm{expt}$
(green line), both show a main peak at 2.0\,eV with a weak shoulder a few tenths of an eV above it, a small peak at
$\sim$4\,eV, and larger one at 5.2\,eV.  $\sigma^\mathrm{BSE}$ aligns very well with $\sigma^\mathrm{expt}$ for the most
part, the main discrepancy in the depth of the trough between 3 and 4\,eV, and the peak at 4\,eV is slightly blue
shifted.  To realize this quality of agreement, it is essential that $\sigma(\omega)$ be calculated from QS$G\hat{W}$.
Optical peaks calculated from the simpler QS$GW$ hamiltonian are blueshifted by $\sim$0.7\,eV corresponding to the blueshift in
the fundamental gap.  Simpler band structures, e.g. LSDA or LSDA+U, bear only a qualitative resemblance to
Fig.~\ref{fig:Fig1}(d)~\cite{Popavic01}.  The optical gap at $\sim$1.8\,eV, is 1\,eV below the fundamental gap.  This
feature is common to many TM antiferromagnetic insulators, such as NiO, \ce{La2CuO4}~\cite{Cunningham23},
CrI$_{3}$~\cite{Acharya22}, and CrSBr~\cite{Shao25}.  To resolve which orbitals govern the optical peaks,
Fig.~\ref{fig:optics}(b) shows a Mulliken decomposition of the joint density-of-states (JDOS), $\sum_{ij\mathbf{k}}
{w^{\mathbf{k}}_iw^{\mathbf{k}}_j\delta(\varepsilon^{\mathbf{k}}_i-\varepsilon^{\mathbf{k}}_j-\omega)}$.  $i$ and $j$
span occupied and unoccupied states, and $w^{\mathbf{k}}_{i}$, $w^{\mathbf{k}}_{j}$ are the Mulliken projection of
normalized eigenstate \textcolor{blue}{with energies} $\varepsilon^{\mathbf{k}}_{i}$, $\varepsilon^{\mathbf{k}}_{j}$.  Some decompositions are given in
Fig.~\ref{fig:optics}(b).  The portion of the total JDOS (black) from pairs of Cu\,$d_{x^2+y^2}$\,+\,O components is
given in red.  Green shows the difference between the two.  Thus only 10\% of JDOS originates from other orbitals up
to $\omega{=}4$\,eV.  It turns out that $w_i$ and $w_j$ are roughly equal, so occupied and unoccupied frontier eigenstates
each have 95\% Cu\,$d_{x^2+y^2}$\,+\,O character.  This character is roughly equally apportioned between Cu and O (not
shown).  The remaining 10\% of JDOS consists almost entirely of $d_{xy}$: blue denotes all contributions other than
$(\mathrm{Cu}\,d_{x^2+y^2}+\mathrm{Cu}\,d_{xy}+\mathrm{O})_\mathrm{occ}\rightarrow(\mathrm{Cu}\,d_{x^2+y^2}+\mathrm{Cu}\,d_{xy}+\mathrm{O})_\mathrm{unocc}$.


Fig.~\ref{fig:optics}(a) also shows $\sigma^\mathrm{BSE}(\omega)$ configurationally averaged from the 20 randomly
disordered structures (dashed violet line).  The main peak is much softened and a new peak at 3.4\,eV appears.  The
relatively poor agreement of configurationally averaged random structures in both optics and photoemission are
consistent with each other and are indicators that short-ranged order is preserved.

Features of the three lowest-energy excitons are depicted in Fig.~\ref{fig:optics}(c), as well as one corresponding to
the shoulder at 2.3\,eV, and a bright exciton at 4.29\,eV.  The minimal unit cell was used for simplicity (the ground
state is similar apart from a $\sim$0.2\,eV blue shift).  The top row of Fig.~\ref{fig:optics}(c) redraws the energy
band structure of Fig.~\ref{fig:Fig1}(d) for each exciton, resolving contributions into one-particle constituents.  The
three lowest-energy excitons are composed almost exclusively from pairs taken from the frontier orbitals, which have
$d_{x^2-y^2}$ character as just noted.  {\color{black} The 1.64, 1.78 and 1.91 eV excitons have negligible onsite $dd$ character and 2.09 eV (shoulder) exciton is the first one that has non-negligible (about 15\%) onsite $dd$ character since spin-allowed transitions between occupied t$_{2g}$ and unoccupied e$_{g}$ states become energetically feasible.} Two-particle states with energies above the fundamental gap is highly delocalized, e.g the exciton shown at 4.29\,eV
with significant Cu $t_{2g}$ character.


\begin{figure}[t]
\centering
\includegraphics[height=4.0cm]{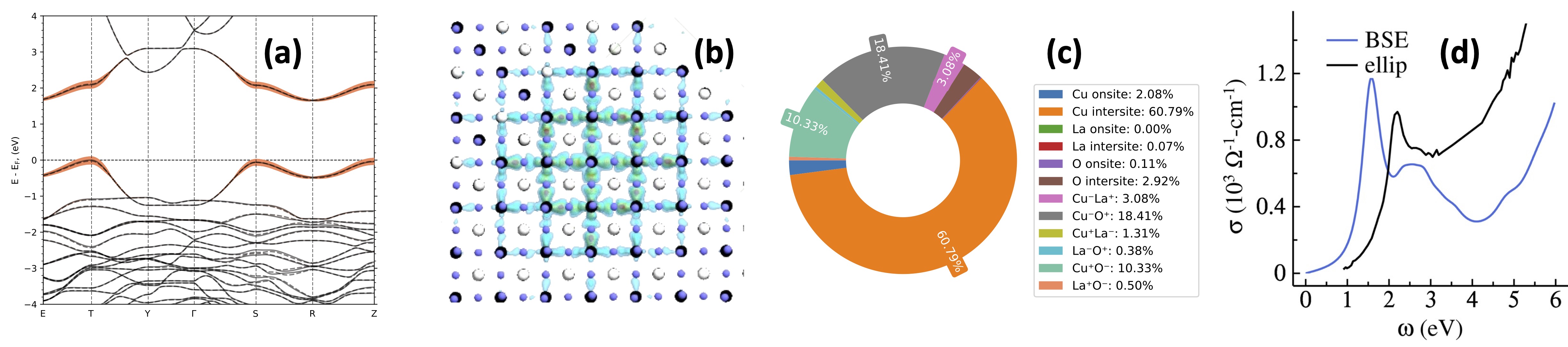}
\caption{Optical properties of \ce{La2CuO4}.
  \textbf{(a)}: QS$G\hat{W}$ band structure of Fig.~\ref{fig:Fig1}(f), projecting the distribution of the pair of lowest-energy excitons into band contributions, as for \ce{SrCuO2} in Fig.\ref{fig:optics} (orange).  
  \textbf{(b)}: Real-space structure of the electron part of the pair of excitons at 1.55\,eV for a hole centered at Cu, projected onto
  in the \textit{xy} plane.  Along \textit{z}  the exciton is confined to a single plane (not shown). Atoms are denoted by spheres: black (Cu), blue (O) and white (La). The exciton constant-value contour is depicted in sky blue.
  \textbf{(c)}: Pie chart resolving the excitons' Mulliken decomposition into atom-atom pairs, distinguishing on-site
  from inter-site contributions. The intersite portion dominates, while the onsite portion is small or negligible.
  \textbf{(d)}: Optical conductivity $\sigma^\mathrm{BSE}(\omega)$ compared to an ellipsometry measurement, Ref.~\cite{Baldini20}.
}
\label{fig:lscooptics}
\end{figure}

The lowest lying excitons in \ce{La2CuO4} bear marked similarities to those in \ce{SrCuO2}, with some differences.
\ce{La2CuO4} has a single frontier orbital each for occupied and unoccupied states as opposed to two; consequently two
optically active subgap excitons appear rather than four, and they are doubly degenerate.  In both systems these
excitons are almost entirely confined to the electron and hole occupying the frontier states (compare
Fig.~\ref{fig:optics}(c) to Fig.~\ref{fig:lscooptics}(a)), which consist of Cu $d_{x^2-y^2}$ admixed with O, and in both
cases the exciton has essentially no extent out-of-plane.  For both systems, the highest-lying bound excitons are composed
mainly of intersite Cu-Cu and Cu-O contributions (compare Fig.~\ref{fig:lscooptics}(c) to Fig.~\ref{fig:optics}(c)).  
Both optical conductivities have a strong peak below the fundamental gap (compare Fig.~\ref{fig:lscooptics}(d) to
Fig.~\ref{fig:optics}(a)).  The peak position agrees less well with ellipsometry measurements in \ce{La2CuO4} than in \ce{SrCuO2},
the optical gap underestimated by 0.4\,eV.  This is a reflection of the liklihood that \ce{La2CuO4} is more strongly
correlated and corrections to the low-order \qsgwh\ diagrams are larger.

For differences, in \ce{La2CuO4} the excitons are symmetric in the plane (Fig.~\ref{fig:lscooptics}(b)), quite
distinct from the anisotropic shape in \ce{SrCuO2} (Fig.~\ref{fig:optics}(c)). The exciton binding energy is much
smaller: $~$0.1\,eV in \ce{La2CuO4} as opposed to $>$0.7\,eV in \ce{SrCuO2}: this is expected owing to
increased screening.  Its linear extent is also similar, spanning about 6 Cu neighbors even though the deeper binding of
\ce{SrCuO2} should reduce its extent.  In this sense, we can view the 1D exciton as ``larger'' one.

\section{Discussions}

The results of  lead us to three primary findings.

1. The canonical picture of spin charge separation in the 1D cuprates is well accounted for in terms of a single Slater determinant, with nonlocal spin disorder being the source of `spinons'.  These appear as an artifact of Unklapp processes originating from the partial destruction of $k$ as a good quantum number, as a result of breaking translational symmetry.

2. \ce{SrCuO2} and \ce{La2CuO4} share a great deal in common: they share a similar ligand structure around the Cu; they form similar local moments; and the bandgap forms for similar reasons --- through formation of bond-antibond pairs that couple $e_g$ states between Cu neighbors.  Both have short range order.  There are also important differences.  \ce{SrCuO2} is highly anisotropic and quasi one-dimensional, as has long been known.  For that reason, it cannot sustain long range order except at extremely low temperatures, unlike \ce{La2CuO4}.  Even so the 2D character is not negligible, and it plays some role in the electronic structure, e.g. modest modification of the bandgap, and contributing in part to the `spinon' band.

3. The lowest lying excitons are rather strongly bound in \ce{SrCuO2} ($\sim$0.7-1\,eV); nevertheless they are far removed from the Frenkel limit and cannot be described by a ligand-field picture as widely assumed {\color{black}where the minimal Hamiltonian for excitons is built from one transition metal ion and its ligands (a CuO$_{4}$ cluster in this case). }
Note that a spin-allowed (non-spin-flip) $d$-$d$ transition can take place in a $d^{9}$ configuration in two possible ways: $(a)$ an onsite transition from $t_{2g}^{6}e_{g}^{3}$ to $t_{2g}^{5}e_{g}^{4}$ and $(b)$ an intersite transition between $d_{x^2-y^2}$ states on neighboring, antiferromagnetically aligned Cu atoms. Since in \ce{SrCuO2} (similarly in \ce{La2CuO4}), the lowest valence and conduction states are exclusively derived from $d_{x^2-y^2}$ character of Cu, the latter option is the only feasible ground state excitonic transition {\color{black}(for all three peaks betwen 1.6 and 1.9 eV)}. In contrast to the picture that emerges from our theory, the ligand-field picture and the associated Sugano-Tanabe diagram~\cite{sugano1954,suganobook} for Cu-$d^{9}$ configuration allows only for option (a), which, as we establish rigorously is not true {\color{black}for the low energy excitons but only for the excited state which appears at the shoulder around 2.3 eV} . In the bottom row of Fig.~\ref{fig:optics}, a Mulliken decomposition of each exciton into atom-atom pairs
is depicted in a {\color{black}horizontal bar plot}.  The largest component of the ground state exciton has intersite Cu-Cu character, with significant Cu-O character as
well, but with negligible on-site Cu $d$-$d$ character.  This highlights the qualitative limitations of the traditional
ligand-field model of electronic structure of these compounds, as is often taken for
granted~\cite{Imada98,Popavic01,Hilgers09}.  The lowest lying excitons are nevertheless fairly short-ranged compared to typical Wannier excitons~\cite{wannier}, and yet sufficiently extended in one direction to counter the traditional Frenkel exciton picture~\cite{frenkelorig1,frenkelorig2,frenkelmolecule, frenkelmolecule2} of a molecular orbital centered around a single Cu atom. {\color{black}Nevertheless, these excitons have strong $dp$ charge transfer character making them ideal candidates for the Zhang-Rice (ZR) picture for excitons as was proposed by Zhang and Rice originally for the two-particle excitations in cuprates~\cite{zhang1988effective} . }

{\color {black}
\section{Conclusions}

We have shown that a self-consistent form of low-order many-body perturbation theory accounts very well for the
``spinon'' and ``holon'' bands, and that the primary signature for spinons (an additional peak in ARPES spectra) is
explained by spin disorder which breaks symmetry and partially destroys $k$ as a good quantum number.  The existing
theory well explains both photoemission and optical experiments without an model or adjustable parameters.
We also show that the system does not appear to be fully disordered: a high degree of short range order is needed
to accurately explain experiments.

Finally, we have pointed out the inadequacy of canonical ligand-field picture used to describe systems with
with Cu in a $d^{9}$ configuration: a multisite, cluster description to account for the bandgap and optical properties.}
{\color{black} In particular, to sufficiently describe the rich excitonic spctrum observed in these materials, as we show from our ab-initio approach that does not make any model approximations, a minimal Hamiltonian that allows for onsite $dd$, intersite $dd$ and charge-transfer Zhang-Rice like $dp$ transitions is absolutely crucial, as the excitonic spectrum crosses over from a intersite $dd$ ($e_{g}$-e$_{g}$)+$dp$ character at low energies to onsite $dd$ ($t_{2g}$-e$_{g}$)+ intersite $dd$  ($e_{g}$-e$_{g}$) + $dp$ at intermediate energies to entirely dipolar intersite $dd$ ($t_{2g}$-e$_{g}$) + $dp$ at high energies.}

\section*{Acknowledgments}

This work was authored by the National Renewable Energy Laboratory, operated by Alliance for Sustainable Energy, LLC,
for the U.S. Department of Energy (DOE) under Contract No. DE-AC36-08 GO28308, funding from Office of Science, Basic
Energy Sciences, Division of Materials.  The U.S. Government retains and the publisher, by accepting the article for
publication, acknowledges that the U.S. Government retains a nonexclusive, paid-up, irrevocable, world-wide license to
publish or reproduce the published form of this work, or allow others to do so, for U.S. Government purposes.  Most of
the large calculations were performed on Frontier computer at Oak Ridge National Lab, under an ALCC grant.  We also made
use of the National Energy Research Scientific Computing Center, under Contract No. DE-AC02-05CH11231 using NERSC award
BES-ERCAP0021783. Finally, a portion of the calculations were performed of Energy’s Office of Energy Efficiency and
Renewable Energy and located at the National Renewable Energy Laboratory.

\bibliography{paper.bib,gw.bib,mvs-paps.bib}

\newpage

\section{Supplemental Information}

Table~\ref{tab:tabs1} provides the lattice vectors and internal coordinates of the minimal unit cell, as depicted in
Fig.~\ref{fig:Fig1}.  The `standard' cell of \ce{SrCuO2}, the presumed ground state, is obtained by doubling the unit
cell, with the change $\mathrm{P}_{1} \rightarrow 2\mathrm{P}_{1} + 2\mathrm{P}_{2}$.  For both \ce{SrCuO2} and
\ce{La2CuO4}, the lattice vectors were rotated to orient Cu's four nearest neighbors north, south, east,
west (See Cu$^{(0)}$ in Fig.~\ref{fig:Fig1}(c)).  This maximizes overlap between the Cu $d_{x^2-y^2}$ orbital and its
nearest neighbors.

\newcommand{\ra}[1]{\renewcommand{\arraystretch}{#1}}
\begin{table*}[h]
\ra{1.2}
\begin{minipage}[t]{.24\textwidth}
\begin{tabular}{@{}rrrr@{}}
\multicolumn{4}{c}{\ce{SrCuO2}} \\
\hline
\vspace{-6pt}\\
P$_1$   & 0.0    & -2.2839 & -0.5  \\
P$_2$   & 0.0    &  2.2839 & -0.5  \\
P$_3$   & 2.1902 &  0.0    &  0.0  \\
\hline
\vspace{-6pt}\\
                 & P$_1$\hspace{12pt} & P$_2$\hspace{12pt} & P$_3$\hspace{6pt} \\ \hline
 Sr              & -0.16904  &  0.16904  &  0.125   \\
 Sr              & -0.16904  &  0.16904  &  0.625   \\
 Sr              &  0.16904  & -0.16904  & -0.125   \\
 Sr              &  0.16904  & -0.16904  &  0.375   \\
 Cu$\uparrow$    &  0.43896  & -0.43896  & -0.125   \\
 Cu$\uparrow$    & -0.43896  &  0.43896  &  0.125   \\
 Cu$\downarrow$  &  0.43896  & -0.43896  &  0.375   \\
 Cu$\downarrow$  & -0.43896  &  0.43896  &  0.625   \\
 O               & -0.32160  &  0.32160  &  0.125   \\
 O               & -0.32160  &  0.32160  &  0.625   \\
 O               & -0.44430  &  0.44430  & -0.125   \\
 O               & -0.44430  &  0.44430  &  0.375   \\
 O               &  0.32160  & -0.32160  & -0.125   \\
 O               &  0.32160  & -0.32160  &  0.375   \\
 O               &  0.44430  & -0.44430  &  0.125   \\
 O               &  0.44430  & -0.44430  &  0.625   \\
\hline
\end{tabular}
\end{minipage}
\hfill
\begin{minipage}[t]{.24\textwidth}
\vskip -142pt
\begin{tabular}{@{}rrrr@{}}
\multicolumn{4}{c}{\ce{La2CuO4}} \\
\hline
\vspace{-6pt}\\
P$_1$   &  0.7071 &  0.7071 & 0.0000  \\                       
P$_2$   & -0.7165 &  0.7165 & 0.0000  \\                       
P$_3$   & -0.3536 & -0.3536 & 1.2448  \\                       
\hline
\vspace{-6pt}\\
                 & P$_1$\hspace{12pt} & P$_2$\hspace{12pt} & P$_3$\hspace{6pt} \\ \hline
 La              & 0.6378  &  0.0066  &  0.2756    \\
 La              & 0.1378  &  0.4934  &  0.2756    \\
 La              & 0.8622  &  0.5066  &  0.7244    \\
 La              & 0.3622  &  0.9934  &  0.7244    \\
 Cu$\uparrow$    & 0.0000  &  0.0000  &  0.0000    \\
 Cu$\downarrow$  & 0.5000  &  0.5000  &  0.0000    \\
 O               & 0.6823  &  0.4699  &  0.3647    \\
 O               & 0.7444  &  0.2500  &  0.9888    \\
 O               & 0.1823  &  0.0301  &  0.3647    \\
 O               & 0.8177  &  0.9699  &  0.6353    \\
 O               & 0.3177  &  0.5301  &  0.6353    \\
 O               & 0.2444  &  0.2500  &  0.9888    \\
 O               & 0.2556  &  0.7500  &  0.0112    \\
 O               & 0.7556  &  0.7500  &  0.0112    \\
\hline
\end{tabular}
\end{minipage}
\caption{Structural data for \ce{SrCuO2} and \ce{La2CuO4}.  Left: minimal cell of \ce{SrCuO2} with 16 atoms.
  Lattice constant was taken to be $a$=3.575\,\AA.
  Right: unit cell of AFM \ce{La2CuO4}.  Lattice constant was taken to be $a$=5.261\,\AA.
  P$_{1}$, P$_{2}$, P$_{3}$ denote the three lattice vectors in units of $a$. Site positions are listed as multiples
  of lattice vectors.
}
\label{tab:tabs1}
\end{table*}

Spin-disordered calculations were performed in 64-atom supercells of the minimal cell, constructed as: $\mathrm{P}_{1}
\rightarrow 2\mathrm{P}_{1} + 2\mathrm{P}_{2}$; $\mathrm{P}_{2}$ unchanged; $\mathrm{P}_{3} \rightarrow
2\mathrm{P}_{3}$.  These are the lattice vectors for the unit cells of Fig.~\ref{fig:Fig2}.

To depict one-particle spectral functions in the minimal cell, energy bands were mapped onto the minimal cell using the
Brillouin zone unfolding technique of Ref.~\cite{TiSe2}.
To model the fully disordered state, 20 configurations of the 64 atom cell were used, with spins randomly configured.
The starting moments were taken from the ordered configuration, and randomly spin flipped.  in all cases the moment
remained quite stable in the course of self-consistency, varying slightly depending on the spin configuration
(Table~\ref{tab:moments}).  This shows that the local moment forms primarily to spin-split the $d_{x^2-y^2}$ orbital;
whether it couples FM or AFM to neighbors is a secondary effect.

\begin{table}[b]
\centering
\ra{1.1}
\begin{tabular}{|@{\hspace{0.6em}}lclll@{\hspace{0.6em}}|}\hline
             & mean  & rms   & min & max\\
\hline
minimal      & 0.58  & 0 && \\
standard     & 0.59  & 0 && \\
config. avg. & 0.64  & 0.04  & 0.57 & 0.72\\
\hline
\end{tabular}
\caption{Self-consistent magnetic moments in $\mu_B$, for the minimal, standard, and configurationally averaged disordered structures.
In the disordered case the mean moment increases relative to the ordered ones, with a small rms fluctuation around the mean.  }
\label{tab:moments}
\end{table}

For the configurationally averaged conductivity of Fig.~\ref{fig:optics}, $\sigma(\omega)$ was calculated for each
configuration separately, and an average taken for the result shown.  For the 1-particle spectral
function the eigenvalues were calculated for each configuration, and the spectral function calculated from a
superposition of the 20 configurations.  The configurational average, the Fermi levels must be aligned.  This was done
by aligning Fermi level determined from Fermi-Dirac statistics at 1000\,K.  (The Fermi level lies near midgap and
changes only slightly with temperature).  All spectral functions shown determined the Fermi level in this way.


\end{document}